\documentclass[preprint,showpacs,aps,prb]{revtex4}
\usepackage[cp1251]{inputenc}
\usepackage[english]{babel}
\usepackage{amsmath}
\usepackage{makeidx}
\usepackage{graphicx}
\usepackage[dvips]{epsfig}
\newcounter{fig}

\begin{document}
\title{Quantum magneto-optics  of graphite family }
\author{L.A. Falkovsky}
\affiliation{ Landau Institute for Theoretical Physics, Moscow
119334\\Verechagin Institute of the High Pressure
Physics, Troitsk 142190}
\date{\today}      

\begin{abstract}
The optical conductivity of graphene, bilayer graphene, and graphite in quantizing magnetic fields is studied.
Both  dynamical  conductivities, longitudinal and Hall's, are 
analytically evaluated. The conductivity peaks are explained in terms of electron transitions. We have shown that  trigonal warping  can be considered within the perturbation theory for  strong magnetic fields larger than 1 T  and in the semiclassical approach for  weak fields when the Fermi energy is much larger than the cyclotron frequency. The main optical transitions obey the selection rule with $\Delta n=1$ for the Landau number $n$, however the $\Delta n=2$ transitions due to the trigonal warping are also possible. 
The Faraday/Kerr rotation and light transmission/reflection in the quantizing magnetic fields are calculated. 
Parameters of the Slonczewski--Weiss--McClure   model  are used in the fit  taking into account the previous dHvA measurements  and correcting some of them for the case of strong magnetic fields.
\end{abstract}
\maketitle
\section{Introduction}
Comprehensive literature on the graphene family can be described in terms of the Dirac gapless fermions. According to this picture, there are two
bands at the $K$  hexagon vertexes of the Brillouin zone  without any gap between them, and the electron dispersion can be considered as linear  in the wide wave-vector region. For the dispersion linearity, this region should be small
compared with the size of the Brillouin zone, i.e. less than 10$^{-8}$ cm$^{-1}$, providing  the small carrier concentration $n\ll 10^{16}$ cm$^{-1}$. Pristine graphene  at  zero temperature has no carriers, and the Fermi level should divide  the conduction and valence bands. However, undoped graphene cannot be really obtained, and so far purest graphene contains about $n\sim10^9$ cm$^{-2}$ of carriers. Then the following problem appears --- how do  Coulomb electron-electron interactions renormalize the linear dispersion and does graphene become an insulator  with a gap?

 Semiconductors with the gap are needed for electronic applications.  
 Investigations of the  graphene bilayer and multilayer are very popular as the  gap appears when  the bias is applied. We see  how physics made a circle for the half of century returning to  graphite studies. Here Slonczewski,  Weiss, and McClure (SWMC) should be mentioned because they 
 have stated the description  of a layered matter \cite{SW} with  interactions strong in the layer and  weak  between layers.
 
  \begin{figure}[]
\resizebox{.5\textwidth}{!}{\includegraphics{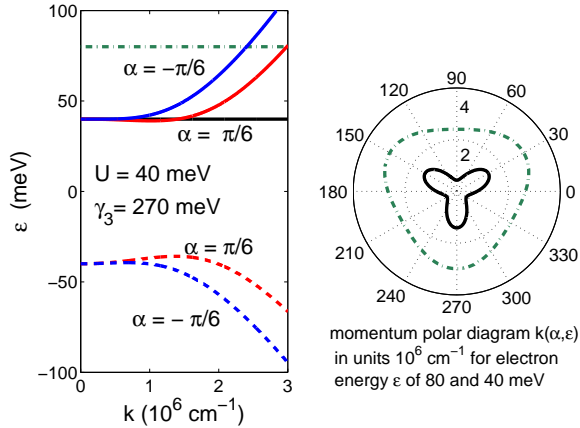}}
\caption{(Color online) (a) The energy dispersion $\varepsilon(k,\alpha)$ of two nearest bands (the electron band shown in solid line and the hole band in dashed line) in bilayer graphene for two polar angles $\alpha$ with the local extrema  at $k\neq 0$ ("mexican hat")  represented. The band  parameters are given in the figure, others are $\gamma_0= 3.05$ eV, $\gamma_1=360$ meV, $ \gamma_4=-150$ meV. (b) Cross-sections $k(\alpha,\varepsilon)$ of the electron band for energies of 80 meV (dashed-dotted line) and 40 meV (solid line).} \label{cr_se}
\end{figure} 
  
  The most accurate  investigation of the band structure of metals and semiconductors is  a study of   the Landau levels  through experiments such as magneto-optics \cite{ST,TDD,DDS,MMD,LTP,OFM,OFS,OP,CLW} and magneto-transport \cite{KTS,LK,JZS,SOP,RM}. In magnetic fields,  the classical and quantum Hall effects are  observed, as well as the polarization rotation for transmitted    
    (the Faraday rotation) or reflected lights (the Kerr rotation). However, the interpretation of the experimental results involves a significant degree of uncertainty, because it is not clear how the resonances
can be identified and which electron transitions they correspond to.
 
  The theoretical  solution for the band problem in magnetic fields  often cannot be  exactly found.  A typical example is presented by  graphene layers. For bilayer graphene and graphite,
 the effective Hamiltonian is a $4\times4$ matrix giving four energy bands.
  Fig. \ref{cr_se}  shows the nearest two bands of the level structure together  with  the semiclassical orbits. 
 The trigonal warping described by the effective Hamiltonian with a relatively small parameter $\gamma_3$ provides an evident effect (see the right panel). Another important parameter is the gate-tunable bandgap $U$ in bilayer graphene. 
  In this situation, the quantization problem cannot be  solved within a rigorous method. To overcome this difficulty several methods have been proposed for approximate \cite{OP,Fa,LA,CBN,ZLB}, numerical \cite{UUU,Nak,PP,GAW,KCD}, and semiclassical quantization \cite{Falko,Dr}. 
  
   The present paper is organized as follows. In Sec. II we recall the electron dispersion in the graphene, bilayer graphene, and graphite. In Sec. III the optical conductivity and light transmission are discussed. In Secs. IV and V  we describe in detail the quantization in magnetic fields. In Sec. VI the longitudinal and Hall conductivities as well as the Faraday/Kerr rotation are described.
  
  \section{Electron dispersion in  graphene family }
\subsection{Electron dispersion in  graphene}
   
The    symmetry  of $K$ point is $C_{3v}$ with the threefold axis and  reflection planes.  This group has twofold  representation with the 
basis functions transforming  each in other under reflections and obtaining the factors $\exp{(\pm2\pi i/3)}$ in rotations. The linear momentum variations from the $K$ point $p_{\pm}=\mp ip_x-p_y$ transform in a similar way. The effective Hamiltonian is invariant under the group transformations, and we have the unique possibility  to construct the invariant  Hamiltonian linear in the momentum as
     \begin{equation}
H(\mathbf{p})=\left(
\begin{array}{cc}
0 \,    & vp_{+} \\
vp_{-} \,& 0      
\end{array}%
\right) \,,  \label{ham0}
\end{equation}%
where $v$ is a constant of the velocity units.  The same Hamiltonian was written using the tight-binding model. 

The eigenvalues of this matrix give  two bands 
\[\varepsilon_{1,2}=\mp v\sqrt{p_x^2+p_y^2}=\mp vp\,,\]
where the sign $\mp$ corresponds to holes and electrons. 
The gapless linear spectrum arises as a consequence of the symmetry, and  the chemical potential at zero temperatures coincides with the band crossing due to the carbon valence. The cyclotron mass has the form
\[m(\varepsilon)=\frac{1}{2\pi}\frac{dS(\varepsilon)}{d\varepsilon}=\frac{\varepsilon}{v^2}\,,\]
and the carrier concentration at zero temperature  
$ n(\mu)=\mu^2/\pi\hbar^2v^2$
is expressed in terms of the chemical potential $\mu$. 

Tuning the gate voltage, the linearity of the spectrum has been examined  in the Schubnikov--de Haas  studies \cite{EGM} with the help of the connection between the effective mass and the carrier concentration  at the Fermi level
$m(\mu)v=\mp\hbar\sqrt{\pi n(\mu)}$.
 The "constant"\, parameter $v$ was found to be no longer constant, but at low carrier concentrations   
$n\sim10^9$ cm$^{-2}$,  it exceeds   its usual value
$v=1.05\pm0.1\times10^8 $ cm/s (at concentrations $n>
10^{11}$ cm$^{-2}$) by the factor of 3.

This is a result of  electron-electron interactions which become stronger at low carrier concentrations. The logarithmic renormalization of the velocity was found by Abrikosov and Beneslavsky in Ref. \cite{AB} for the 3d case
and in Refs. \cite{Mi,GGV} for 2d graphene. Notice, that no phase transition  was revealed even at lowest carrier concentration. We can conclude that  the Coulomb interactions do not create any gap  in the spectrum.


\subsection{Electron dispersion in  bilayer graphene and graphite }

  Bilayer graphene has attracted much interest partly due to the
opening of a tunable gap in its electronic spectrum by  an
external electrostatic field. Such a phenomenon was predicted in
Refs.  \cite{McF,LCH} and was observed in optical studies
controlled  by applying a gate voltage
\cite{OBS,ZBF,KHM,LHJ,ECNM,NC,MLS,KCM}. 
 
\begin{figure}[h]
\resizebox{.35\textwidth}{!}{\includegraphics{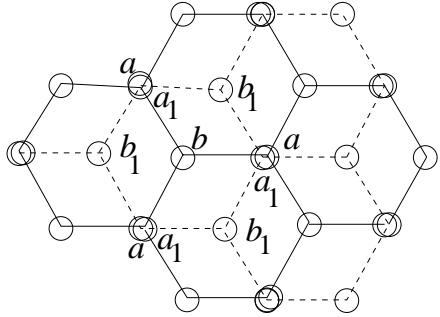}}
\caption{Bilayer lattice} \label{grlat}
\end{figure}

 The  graphene bilayer lattice is shown in Fig. \ref{grlat}.
Atoms  in one layer, i.\,e., $\bf{a}$ and $\bf{b}$ in the unit cell, are
connected by  solid lines, and in the other layer, e.\,g., $\bf{a_1}$
and $\bf{b_1}$, by the dashed lines. The atom $\bf{a}$ ($\bf{a_1}$) differs from
$\bf{b}$ ($\bf{b_1}$) because it has a neighbor  in the adjacent
layer, whereas the atom $\bf{b}$ ($\bf{b_1}$) does not.

  The effective Hamiltonian of the SWMC theory can be written \cite{PP,GAW} near the $ KH$ line  in graphite as
\begin{equation}
H(\mathbf{p})=\left(
\begin{array}{cccc}
\tilde{\gamma}_5     & vp_{+} & \tilde{\gamma}_1 &\tilde{\gamma}_4vp_{-}/\gamma_0\\
vp_{-} & \tilde{\gamma}_2  & \tilde{\gamma}_4vp_{-}/\gamma_0& \tilde{\gamma}_3vp_{+}/\gamma_0\\
\tilde{\gamma}_1   &\tilde{\gamma}_4vp_{+}/\gamma_0 & \tilde{\gamma}_5   &vp_{-}\\
\tilde{\gamma}_4vp_{+}/\gamma_0 & \tilde{\gamma}_3vp_{-}/\gamma_0 &vp_{+} &\tilde{\gamma}_2
\end{array}%
\right)\,,   \label{hamg}
\end{equation}%
where $p_{\pm}=\mp ip_x-p_y$ are the momentum components and 
$\tilde{\gamma}_j$ are the functions of the  $p_z$ momentum in the major axis direction,
\begin{eqnarray}\tilde{\gamma_2}=2\gamma_2\cos{(2p_zd_0)}\,,\tilde{\gamma}_5=2\gamma_5\cos{(2p_zd_0)}+\Delta\,,\nonumber\\ 
\tilde{\gamma}_i=2\gamma_i\cos{(p_zd_0)}\quad\text{for}\quad i=1,3,4,\nonumber  
\end{eqnarray}
with the distance $d_0=3.35$ \AA\, between   layers in graphite. 
 The nearest-neighbor hopping integral $\gamma_0\approx 3$ eV corresponds with the velocity parameter $v=1.5a_0\gamma_0 = 10^6$ m/s and the in-layer inter atomic distance $a_0=1.415$ \AA\,. 
The Hamiltonian  (\ref{hamg})  is represented in a somewhat different form than that in Ref. \cite{SW}. The relations between the hopping integrals in these forms are given in Table \ref{tb1}.  The recent
estimate \cite{KHM,LHJ} of the parameters agrees with those given in the Table.

\begin{table}[]
\caption{\label{tb1} The parameters of the Hamiltonian, Eq. (\ref{hamg}), their values in the SWMC model, and obtained in the experimental works, all in meV.  }
        \begin{ruledtabular}
                \begin{tabular}{cccccccccc}
 &Eq. (\ref{hamg})&$\gamma_0$ & $\gamma_1$ & $\gamma_2$& $\gamma_3$ & $\gamma_4$ & $\gamma_5$ & $\Delta$ & $\varepsilon_F$\\
 & &3050& 360&$-10.2$&270&$-150$&$-1.5$&16&$-4.1$\\
 \hline
 &S$^a$& $\gamma_0$ & $\gamma_1$ & $2\gamma_2$& $\gamma_3$ & 
 $-\gamma_4$ & $2\gamma_5$ & $\Delta+2(\gamma_2-\gamma_5)$&  2$\gamma_2+\varepsilon_F$\\  
&M$^b$& 3160& 390&$-20$&276&44&38&8&$-24$\\ 
&D$^c$& 3120& 380&$-21$&315&120&$-3$&$-2$&$-$\\ 
&DFT$^d$& 2598$\pm$15&340$\pm20$&- &320$\pm$20&177$\pm$25&- &24$\pm$10&-
\end{tabular}
\end{ruledtabular}
 $^a$SWMC,Ref. \cite{SW},
 $^b$Mendez et al, Ref. \cite{MMD},
 $^c$Doezema et al, Ref. \cite{DDS}.
 $^d$Charlier et al, Ref. \cite{CM}.
\end{table} 

 \begin{figure}[h]
\resizebox{.4\textwidth}{!}{\includegraphics{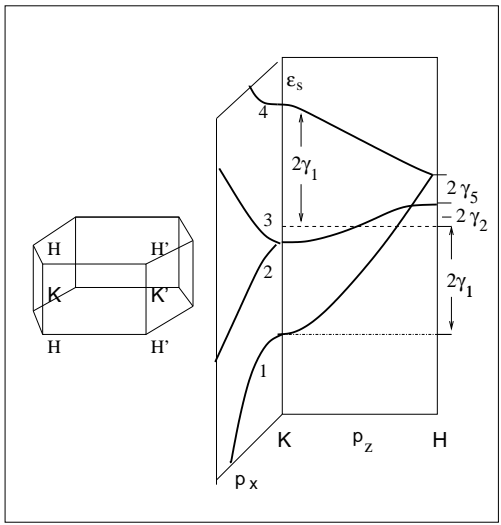}}
\caption{Band structure of graphite} \label{bandg}
\end{figure}
 
The electron spectrum  of graphite is shown in Fig. \ref{bandg}. There are four levels labeled by the number $s=1,2,3,4$ from below at any
momentum. As a consequence of axial symmetry, twofold degeneration  $\varepsilon_2=\varepsilon_3$ exists at  $p_x=p_y=0$, i. e., on the $ KH$ line.

In bilayer graphene, every layer has only one neighboring layer. Therefore, we have to set
$\gamma_2=\gamma_5=0$ and to substitute $\tilde{\gamma}_i=\gamma_i$  for $i=1,3,4$ in Hamiltonian (\ref{hamg}).
The parameter $U$ can be included also in the bilayer Hamiltonian as a result of the gate voltage. 
Then, the gap appears between the $\varepsilon_2$ and  $\varepsilon_3$, and these bands acquire the form of "mexican hat"\,. An important point is that two points, $K$ and $K'$, are in the Brillouin zone transforming each in other under reflection. Such a reflection changes the $U$ sign giving two different dispersion laws at the $K$ and $K'$ points. 
 
\section{Optical conductivity}

We use the general expression for the conductivity as a function of  the electric field frequency $\omega$ and wave vector $k$ in the form  \cite{FV,GSC}
\begin{eqnarray}  \label{con}
&&\sigma _{ij}(\omega ,k) =\nonumber\\
&& 2ie^{2}\sum_{\mathbf{p},m>n}\left\{ \frac{v_{mm}^{i}v_{mm}^{j}\{f_0[\varepsilon _{m}(%
\mathbf{p}_{-})]-f_0[\varepsilon
_{m}(\mathbf{p}_{+})]\}}{[\varepsilon
_{m}(\mathbf{p}_{+})-\varepsilon _{m}(\mathbf{p}_{-})][\omega
-\varepsilon
_{m}(\mathbf{p}_{+})+\varepsilon _{m}(\mathbf{p}_{-})]}\right.   \\
&&\left. +2\omega  \frac{v_{mn}^{i}v_{nm}^{j}\{f_0[\varepsilon
_{m}(\mathbf{p}_{-})]-f_0[\varepsilon _{n}(\mathbf{p}_{+})]\}}{%
[\varepsilon _{n}(\mathbf{p}_{+})-\varepsilon
_{m}(\mathbf{p}_{-})]\{(\omega+i\delta)
^{2}-[\varepsilon _{n}(\mathbf{p}_{+})-\varepsilon _{m}(\mathbf{p}%
_{-})]^{2}\}}\right\}\,,  \nonumber
\end{eqnarray}
 valid in the collisionless limit ($\omega, kv)\gg \tau^{-1}$, where $\tau^{-1}$ is the  electron relaxation frequency,  
 ${\bf p}_{\pm}={\bf p}\pm{\bf k}/2$, and $v^i_{mn}$ is the matrix element of the velocity operator 
  \begin{equation}
  \mathbf{v}=\partial
H(\mathbf{p})/\partial \mathbf{p}
\label{vel}\end{equation}
   determined by  Hamiltonians (\ref{ham0}) or (\ref{hamg}).
Hitherto, we did not use any peculiarities of the graphene spectrum
The expression acquired only the factor 4 due
to summation over spin and over six points of the $K$ type (two
per the Brillouin zone).

The first term in Eq. (\ref{con}) corresponds to the intraband
electron-photon scattering processes. In the limit of the high
carriers concentration $ (T,E_{F})\gg kv$, it coincides with
the usual Drude--Boltzmann conductivity, if the substitution $\omega\rightarrow  \omega+i\tau^{-1}$ is made. 
The second term  owes its origin to the
interband $ n\rightarrow m$
transitions with the infinitesimal    $\delta$ determining the bypass around the pole while integrating over the momentum  ${\bf p}$. The real part of this contribution  is reduced to
the well-known expression for the absorbed energy due to  direct
interband transitions. 
  
\subsection{Optical conductivity of graphene}
For  optical frequencies $\omega\gg kv$,  we can integrate in Eq. (\ref{con}) over the angle and write the conductivity as 
\begin{equation}
  \sigma(\omega) = \frac{e^2\omega}{i\pi\hbar}\left
  [\int\limits_{-\infty}^{+\infty} d\varepsilon\frac{|\varepsilon |}{\omega ^2}
   \frac{df (\varepsilon)}{d\varepsilon}- \int\limits_0^{+\infty}
  d\varepsilon\frac{f (-\varepsilon)-f(\varepsilon)}{(\omega+i\delta)^2 -
  4\varepsilon^2}\right]\, 
  \label{sigma}
\end{equation}
using the variable $\varepsilon=vp$.

The  intraband term can be integrated once more,
\begin{equation}
     \sigma ^{intra}(\omega) =\frac{2ie^2T}
     {\pi\hbar(\omega+i\tau^{-1})}
\ln{(2\cosh\frac{\mu}{2T})}
 \label{sigm}    \, ,
 \end{equation}
where we write $\omega +i\tau^{-1} $ instead of $\omega $  to take  the small relaxation frequency into account.  This Drude--Boltzmann conductivity at low temperatures $T\ll\mu$ takes the form 
\begin{equation}
     \sigma ^{intra}(\omega) =\frac{ie^2|\mu|}
     {\pi\hbar(\omega+i\tau^{-1})}
 \label{si}    \, .
 \end{equation}
 In the opposite limit of high temperatures, the intraband conductivity Eq. (\ref{sigm}) becomes 
 \begin{equation}
     \sigma ^{intra}(\omega) =\frac{2ie^2T\ln{2}}
     {\pi\hbar(\omega+i\tau^{-1})}
\label{sigm1}    \, .
 \end{equation}
 The temperature dependence of the relaxation rate in graphene is discussed theoretically in Ref. \cite{Fatem}.

The interband contribution in Eq.  (\ref{sigma}) integrated at zero temperatures gives
\begin{equation}
    \sigma^{inter}(\omega) =
    \frac{e^2}{4\hbar}\left[\theta(\omega-2\mu)-\frac{i}{2\pi}\ln
    \frac{(\omega+2\mu)^2}{(\omega-2\mu)^2}
     \right]\,,
 \label{ibd} \end{equation}
where the $\theta-$function expresses the  threshold behavior of interband electron transitions at $\omega=2\mu$. The temperature  smooths out all the singularities
\begin{eqnarray}
\theta(\omega-2\mu)\rightarrow\frac{1}{2}+\frac{1}{\pi}\arctan\frac{\omega-2\mu}{2T}\\
\nonumber (\omega-2\mu)^2\rightarrow(\omega-2\mu)^2+(2T)^2\,.
\label{sub}\end{eqnarray}

The main issue should be emphasized. In high frequency region
$\omega\gg(T,\mu)$, the iterband transitions make the main contribution into  conductivity 
$$\sigma(\omega) =
    \frac{e^2}{4\hbar},$$
     having the universal character independent of any material parameters. This frequency region is limited above by the band width of around 3 eV. Making use the universal conductivity, one can calculate the light transmission through graphene \cite{FP} in the  approximation  linear in conductivity 
\begin{equation}
  T=1-\frac{4\pi}{c} Re\,\sigma(\omega)\cos{\theta}=1-\pi\frac{e^2}{\hbar
  c}\cos{\theta}\,,
  \label{transmis}\end{equation}
where $\theta$ is the incidence angle. The intensity of reflected light is quadratic in the fine structure constant $\alpha= e^2/\hbar c $. In excellent agreement with the theory, for the wide optical range,   several experimental groups \cite{NBG, Li, Ma} observe   the light transmission through graphene as well as bilayer graphene where the difference from unity is twice as larger.  It is exceptionally intriguing that the light transmission involves the fine structure constant of quantum electrodynamics having really no relations to the graphene physics. 

For graphite, the value of $\sigma_d=e^2/4\hbar d_0$ plays the role of the universal dynamical conductivity, where $d_0$ is the distance between layers. As shown experimentally \cite{KHC} and theoretically \cite{Fatem}, the dynamical conductivity of graphite is close to this universal value in the frequency range 0.1--1 eV having the kink singularity at the  interband transition frequency $\omega=2\gamma_1$.


\section{Graphene in magnetic fields}
In the presence of the magnetic field $B$, the momentum projections $p_+$ and $p_-$ become the operators with the commutation rule $\{\hat{p}_{+},\hat{p}_{-}\}=-2e\hbar B /c$. 
 We  use the relations
\[v\hat{p}_+=\omega_B\,a, \quad  v\hat{p}_-=\omega_B\,a^+\]
involving the creation $a^+$ and annihilation $ a$ operators  with  $\omega_B=v\sqrt{2|e|\hbar B/c}$\,  . We will write only one of two $x,y$ space coordinates including the corresponding degeneracy proportional to the magnetic field in the final results.

  For graphene, we search the eigenfunction of Hamiltonian  (\ref{ham0}) in the form
\begin{equation}
\psi_{sn}^{\alpha}(x)=
\left\{\begin{array}{c}
 C^{1}_{sn}\varphi_{n-1}(x)\\
 C^{2}_{sn}\varphi_{n}(x)\,
\end{array}\right.\,,\label{funcg}
\end{equation}
where   $\varphi_{n}(x)$ are 
orthonormal Hermitian   functions with the Landau number $n\ge0$. Canceling the Hermitian functions from the equations, we obtain a system of  linear equations for the eigenvector ${\bf C}_{sn}$ 
\begin{equation}
\left(
\begin{array}{cc}
-\varepsilon     & \omega_B\sqrt{n}\\
\omega_B\sqrt{n} & -\varepsilon     
\end{array}%
\right) \times\left\{\begin{array}{c}
C^{1}_{sn}\\
C^{2}_{sn}
\end{array} \right.=0\,  \label{ham1}
\end{equation}
giving the  eigenvalues
 \[\varepsilon_{sn}=\mp \omega_B\sqrt{ n}\,\] 
 with  $s=1,2$   and $ n=0,1, 2...$
For $n=0$, there is only one level $\varepsilon_{10}=0$ with $C^1_0=0, C^2_0=1 $ as follows from Eq. (\ref{funcg}).
The wave function columns write
\begin{equation}
\begin{array}{c}
C^{1}_{sn}\\
C^{2}_{sn}
\end{array} = \frac{1}{\sqrt{2}}\left\{\begin{array}{c}\, 
1 \\
-1  \end{array}\quad \text{and}\quad \begin{array}{c} 1\\1 \end{array}\,\right.\label{ham2}
\end{equation}
for  $s=1$ and  $s=2$ and  $n=1,2...$.

\section{Graphene layers with trigonal warping in magnetic fields}
 We search the eigenfunction of Hamiltonian  (\ref{hamg}) as a column 
\begin{equation}
\psi_{sn}^{\alpha}(x)=
\left\{\begin{array}{c}
 C^{1}_{sn}\varphi_{n-1}(x)\\
 C^{2}_{sn}\varphi_{n}(x)\\
 C^{3}_{sn}\varphi_{n-1}(x)\\
 C^{4}_{sn}\varphi_{n-2}(x)\,
\end{array}\right.\,.\label{func}
\end{equation}

 One sees the every row
in  Hamiltonian (\ref{hamg}) becomes proportional to the definite Hermitian function if the terms with $\gamma_3$ are omitted. We will show that the terms proportional to $\gamma_3/\gamma_0$ can  be   considered  within the perturbation theory or the semiclassical approximation.

Canceling  the Hermitian functions from the equations,  we obtain 
a system of the linear equations for the eigenvector~${\bf C}_{sn}$
\begin{equation}
\left(
\begin{array}{cccc}
\tilde{\gamma}_5-\varepsilon     & \omega_B\sqrt{n} & \tilde{\gamma}_1  & 
\omega_4\sqrt{n-1}\\
\omega_B\sqrt{n} & \tilde{\gamma}_2-\varepsilon      & \omega_4\sqrt{n}& 0\\
\tilde{\gamma}_1      &\omega_4\sqrt{n} & \tilde{\gamma}_5-\varepsilon   &\omega_B\sqrt{n-1}\\
\omega_4\sqrt{n-1}& 0 &\omega_B\sqrt{n-1} &\tilde{\gamma}_2-\varepsilon
\end{array}%
\right) \times\left\{\begin{array}{c}
C^{1}_{sn}\\
C^{2}_{sn}\\
C^{3}_{sn}\\
C^{4}_{sn}
\end{array} \right.=0\, , \label{ham2}
\end{equation}
where the band number $s=1,2,3,4$ numerates the solutions at given $n$  from the bottom, $\omega_B=v\sqrt{2|e|\hbar B/c}$\,  and\, $\omega_{4}=\tilde{\gamma}_{4}\omega_B/\gamma_0$.
 
\begin{figure}[]
\resizebox{.52\textwidth}{!}{\includegraphics{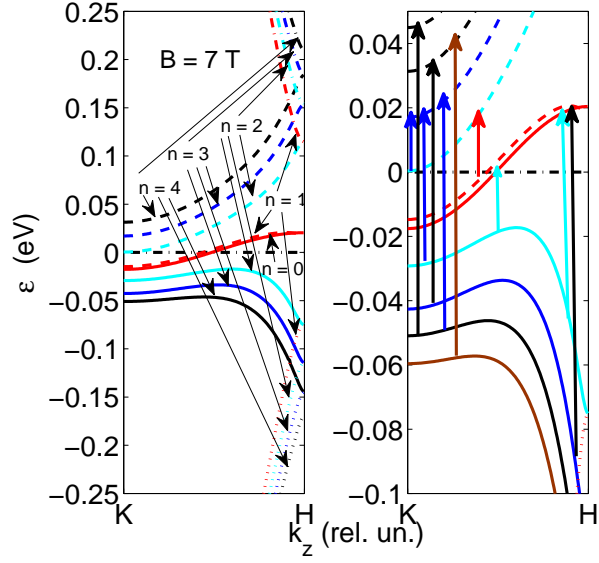}}
\caption{(Color online) Landau levels in graphite $\varepsilon_{sn}$ for $n$ from 0 to 4 in four bands $s=$1,2,3, and 4 (in dotted, solid, dashed, and dash-dotted lines, correspondingly) as functions of wave vector $k_z$ along the $KH$ line in the Brillouin zone 
($K=0,\, H = \pi/2d_0$) in  the magnetic field $B$ = 7~T with the SWMC model parameters given in Table 1.
  The main electron transitions shown in the right panel below 100 meV occur between the levels with the selection rule  $\Delta n = 1$\,, see text. }
\label{d7strf}\end{figure}

The eigenvalues of the matrix in Eq. (\ref{ham2}) are easily found,  they are shown in Fig. \ref{d7strf} as a function of the momentum $p_z$. For each Landau number  $n\ge 2$ and momentum $p_z$, there are four eigenvalues
 $\varepsilon_{s}(n)$  and four corresponding eigenvectors, Eq.  (\ref{func}), marked by  the band subscript  $s$.    We  use the notation $|sn\rangle$ for levels.
  In addition, there are four  levels. One of them is \begin{equation}\varepsilon_1(n=0)=\tilde{\gamma}_2\label{n0}\end{equation}
 for $n=0$ with the eigenvector ${\bf C}_0=(0,1,0,0)$ as is evident from Eq. (\ref{func}). It  intersects   the Fermi level
 and belongs to the  electron (hole) band near the $K$     $(H)$ point. The other  three levels  indicated with $n=1$  and $s=1,2,3$ are determined by the first three equations of the system (\ref{ham2}) with $C^4_{s1}=0$.

  The $|21\rangle$ level is  close to the $|10\rangle$ level. In the  region of $p_z$,  $\gamma_1/\cos{2p_zd_0}\gg \gamma_2$, where the electrons are located, this level has the energy 
\[\varepsilon_2(n=1)=\tilde{\gamma}_2-2\frac{\omega_B^2\tilde{\gamma}_4}{\tilde{\gamma}_1\gamma_0}\,.\]
 In the same region, the two closest bands ($s=2,3$) with 
 $n\ge 2$ are written as
 \begin {equation}
\begin{array}{c}
\varepsilon_{2,3}(n)={\displaystyle\tilde{\gamma}_2-\frac{\omega_B^2\tilde{\gamma}_4}{\tilde{\gamma}_1\gamma_0}(2n-1)} 
{\displaystyle\mp
\frac{\omega_B^2}{\tilde{\gamma}_1}\sqrt{n(n-1)}}\,
\end{array}\label{de1}\end{equation}
 within accuracy of $(\tilde{\gamma}_4/\gamma_0)^2$.
 
 \subsection{Perturbation theory for matrix Hamiltonian}
 Due to a double degeneracy existing on the $KH$ line, the effect of the trigonal warping becomes essential.  
 A simplest way \cite{Fa} to evaluate the corrections  resulting from the warping $\gamma_3$ consists in the consideration of the  Green's function having the poles at the electron levels.
 
  The Green's function of the unperturbed  Hamiltonian  writes using the functions in Eq. (\ref{func}) as
 \begin{equation}
 G^{\alpha \beta}_0(\varepsilon,x,x')=\sum_{sn}\frac{\psi^{\alpha}_{sn}(x) \psi^{*\beta }_{sn}(x')}{\varepsilon-\varepsilon_{sn}}\,.\label{gf}\end{equation}
 The corrections to the levels can be found in the iterations  
 \begin{equation}{\bf G}_{m+1}(x,x')=\int d^2x''{\bf G}_0(x,x''){\bf V}(x''){\bf G}_m(x'',x')\,,\label{it}\end{equation}
where $\bf V(x)$ has only two matrix elements $V^{42}=\omega_B\tilde{\gamma_3}a^{+}/\gamma_0$ and $V^{24}=V^{42*}$ in the  Hamiltonian (\ref{hamg}).

\begin{figure}[]
\resizebox{.25\textwidth}{!}{\includegraphics{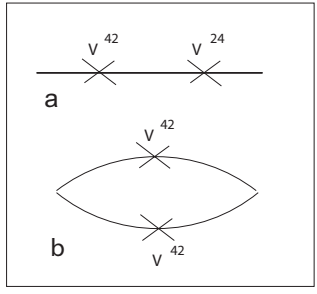}}
\caption{ Diagrams for the second iteration of the perturbation theory; corrections to the Green's function (a), corrections to the vertex in  conductivity (b).}
\label{diag}\end{figure}

In the second iteration, we 
get the corrections
\[\int d^2x_1 d^2x_2G_0^{\alpha4}(x,x_1)V^{42}(x_1)G_0^{22}(x_1,x_2)V^{24}(x_2)G_0^{4\beta}(x_2,x')\]
and the similar term with the superscript  substitution  $2\leftrightarrow4$. The matrix elements of the perturbation $V$ are easily calculated with respect to the Hermitian functions in Eqs. (\ref{gf}) and (\ref{func}), and we obtain  
\begin{equation}\left(\frac{\omega_B\tilde{\gamma}_3}{\gamma_0}\right)^2\sum_{s'sn}\frac{(n-2)|C^4_{sn}C^2_{s',n-3}|^2\psi^{\alpha}_{sn}(x) \psi^{*\beta }_{sn}(x')}{(\varepsilon-\varepsilon_{sn})(\varepsilon-\varepsilon_{s',n-3})(\varepsilon-\varepsilon_{sn})}.\label{corr1}\end{equation}
for the diagram shown in the upper part of Fig. \ref{diag}.
This correction plays an important role near the poles of the Green's
function. For this reason, for  $\varepsilon$  close to $\varepsilon_{sn}$, the $\varepsilon$ value in the second factor of the denominator can be replaced by $\varepsilon_{sn}$. Thus, the total Green's function (with the correction) has the structure
\[\frac{1}{\varepsilon-\varepsilon_{sn}}+\frac{\delta}{(\varepsilon-\varepsilon_{sn})^2}\,,\] 
which  can be rewritten  to the second-order terms in $\delta$ as
\[\frac{1}{\varepsilon-\varepsilon_{sn}-\delta}\,.\]
Therefore, we can  represent the correction as a shift $\delta\varepsilon_{sn}$ of the poles $(\varepsilon-\varepsilon_{sn}-\delta\varepsilon_{sn})^{-1}$ with
 \begin{equation}
 \begin{array}{c}
{\displaystyle\delta\varepsilon_s(n)=\left(\frac{\omega_B\tilde{\gamma}_3}{\gamma_0}\right)^2\sum\limits_{s'}\left\{\frac{(n-2)|C^4_{sn}C^2_{s',n-3}|^2}{\varepsilon_{s}(n)-\varepsilon_{s'}(n-3)}\right.}\\ 
+{\displaystyle\left.\frac{(n+1)|C^2_{sn}C^4_{s',n+3}|^2}{\varepsilon_{s}(n)-\varepsilon_{s'}(n+3)}\right\} }\,,
\end{array}
\label{pt1}\end{equation}
where the first term should be omitted for $n-3<0$. 
In fact, our illustration is nothing but a calculation of the electron self-energy and the naive expansion of the denominator can  indeed be replaced by  summarizing of the corresponding diagrams.

 The corrected $|10\rangle$ level  writes \begin{equation}\varepsilon_1(n=0)=\tilde{\gamma}_2+\left(\frac{\omega_B\tilde{\gamma}_3}{\gamma_0}\right)^2\sum\limits_{s'}\frac{|C^4_{s'3}|^2}
{\tilde{\gamma}_2-\varepsilon_{s'}(3)}\,.\label{nc0}\end{equation}   
  The $|21\rangle-$level 
  is very close to the level with $n=0$, Eq. (\ref{nc0}). 

Comparing
the corrections, Eq. (\ref{pt1}), with the main contribution Eq. (\ref{de1}), we find, first, that the
perturbation theory is valid  when an  expansion parameter $(\tilde{\gamma}_3\tilde{\gamma}_1/\gamma_0\omega_B)^2$ becomes small, i.e., for  strong magnetic fields $B> 1\, T$. Second, the effect of $\gamma_4$ is linear, whereas 
of $\gamma_3$ is quadratic in these constants. Therefore, the $\gamma_4$ constant is more essential for the electron levels in  magnetic fields.

  Comparison shows that Eqs. (\ref{pt1}) and (\ref{nc0}) for levels give the same results as the numerical method of truncating the infinite-rink matrix in Ref. \cite{Nak}.

Note that the derived expressions  are applicable as well to  bilayer graphene while one includes  the field $U$ and substitutes  $\gamma_2=\gamma_5=0$ and $\tilde{\gamma_i}=\gamma_i$ for $i=1,3,4$. In the simplest approach, when only main parameters $\gamma_1$ and $U$  
are holded, the magnetic levels $\varepsilon_{sn}$ are determined by the equation
\[[(U-\varepsilon_{sn})^2- \omega_B^2n][(U+\varepsilon_{sn})^2- \omega_B^2(n+1)]+\gamma_1^2(U^2-\varepsilon_{sn}^2)=0\,.\]


\subsection{ Berry phase, semiclassical quantization  and Landau levels}
Alternatively,  the semiclassical quantization can be applied for relatively weak magnetic fields when the cyclotron frequency is small compared to the Fermi energy.
 Then, we can use the Bohr--Zommerfeld condition as
\begin{equation}
\frac{c}{e\hbar B}S(\varepsilon)= 2\pi\left[n_c+\frac{\mathcal T}{4}+\delta(\varepsilon)\right]\,.
\label{on}\end{equation}
Here $S(\varepsilon)$ is the cross-section area of the electron orbit in the $p_x. p_y$ space for the energy $\varepsilon$  and  the constant momentum projection $p_z$   on the magnetic field, $n_c$ is an integer supposed to be large. The integer $\mathcal T$ is the number of the smooth turning points on the electron orbit. There are two smooth turning  points for the Landau levels and only one for  skipping electrons reflected by the hard edge.

We use the semiclassical approach  for the  magnetic field normal to the layered system when 
the in-layer  momentum components $p_x$ and $p_y$ are only quantized and  the size of the Fermi surface is small  compared with the Brillouin zone size. 
Notice, that the  $\delta(\varepsilon)$-phase depends on the energy. If  the spin is neglected,  $\delta=0$ and $\mathcal T=2$ for the Landau levels, and $\delta=1/2$ and $\mathcal T=2$ for
monolayer graphene. In these two cases, the semiclassical  result
coincides  with the rigorous quantization and it is closely connected
with the topological Berry phase \cite{Be}.
 This $\delta$-phase was evaluated for bismuth in  Ref. \cite{Falko}, preceding Berry's work by almost two decades, and it was considered again for bismuth in Ref. \cite{MS}. For graphite,  the semiclassical quantization was applied in Ref. \cite{Dr}. However, in the general case, the evaluation of the $\delta-$phase is still attracted a widespread interest \cite{TA,CU,KEM,PM,PS,LBM,ZFA}.

 The problem under consideration is described  by the
Hamiltonian in Eqs. (\ref{ham0}) or (\ref{hamg}) rewritten in the form
\begin{equation}\label{hami}
({\bf V\cdot\tilde{p}} +\Gamma-\varepsilon)\Psi=0\,, \end{equation}
where ${\bf \tilde{p}}$ and ${\bf V}$ are the two-dimansional vector and matrix, correspondingly, with the in-layer components $x$ and $y$. The column $\Psi$ is labeled by the band subscript
which we omit together with the matrix subscripts on   $\Gamma$ and ${\bf V}$,  summation over them is implied in Eq. (\ref{hami}). Matrices
 $\Gamma$ and ${\bf V}$ are the first two terms (of  zero and first orders) in a series expansion of the Hamiltonian in the power of quasi-momentum $p_x$ and $ p_y$. 
 
 In the magnetic field, the momentum operator ${\bf\tilde{p}}$
depends on the vector-potential ${\bf A}$  by means of the Peierls substitution, 
$${\bf\tilde{p}}=-i\hbar\nabla-e{\bf A}/c,$$ providing the gauge invariance of the theory. 
The magnetic field can also enter 
  explicitly describing the magnetic interaction with a  spin of
 particles. However, for the graphene family, the magnetic interaction is weak and  omitted here.

It is convenient to choose the
vector-potential in the Landau gauge  $A_x=-By, A_y=A_z=0$ 
in such a way that the Hamiltonian does not depend  on the $x$ coordinate. 
We search the function $\Psi$ in the form
\begin{equation}\nonumber
\Psi=\Phi \exp{(is/\hbar)}\,,
\label{be}\end{equation}
where the function $s$ is assumed to be common for all component
of the column $\Psi$.

The function $\Phi$ is expanding in series of $\hbar/i$:
\[
\Phi=\sum_{m=0}^{\infty}\left(\frac{\hbar}{i}\right)^m
\varphi_m\,.
\]
Collecting the terms with the same powers
of $\hbar$ in Eq. (\ref{hami}), we have
\begin{equation} \label{ham3}({\bf V\cdot p}
+\Gamma-\varepsilon)\varphi_m=-{\bf V \nabla}\varphi_{m-1}\,.
\end{equation}
For $m=0$, we get a homogeneous system of algebraic equations for the wave function column $\varphi_0$,
\begin{equation} \label{ham4}({\bf V\cdot p}
+\Gamma-\varepsilon)\varphi_0=0\,,
\end{equation}
which has a solution under the condition
\begin{equation} \text{Det}({\bf V\cdot  p}
+\Gamma-\varepsilon)=0\,.
\label{edm}\end{equation}
This equation determines  the classical electron orbit,
$\varepsilon(p_x,p_y)=\varepsilon$, at  the given electron energy  $\varepsilon$  in presence of the magnetic field.
On the other hand, the equation coincides  with 
 the  dispersion  equation since it does not contain the magnetic field. In 3d
case, as in graphite, the  dispersion depends also on the momentum
projection $p_z$ on the magnetic field. Thus, our scheme does not requires the expansion in a power of $p_z$.

The equations (\ref{ham3}) with $m=0, 1$ give the wave function in the semiclassical approximation \cite{Falko}. The quantization condition can be written as usual from the requirement that the wave function has to be single-valued. 
  Making the bypass in the complex plane around the turning points 
  to obtain the decreasing solutions in the classically unaccessible region, we obtain, first,   $\mathcal{T}=2$ and, second, $\delta$-phase  as a contour integral along the classical orbit 
\begin{equation} \label{on1}
\delta(\varepsilon)=\frac{1}{2\pi}\text{Im}
\oint\frac{dp_x}{\varphi^*_0\varphi_0 v_y}
\varphi^*_0V_y\frac{d\varphi_0}{dp_x}\,,
\end{equation}
where $v_y=\partial \varepsilon(p_x,p_y)/\partial p_y$.
 Using the Hamiltonian hermiticity, after the simple
algebra (see Ref. \cite{Falko}), Eq. (\ref{on1}) can be rewrite in the gauge-invariant form
\begin{equation} \label{on2}\delta(\varepsilon)=\frac{1}{4\pi}\text{Im}
\oint\frac{dp}{\varphi_0^*\varphi_0 v} \varphi^*_0\left[{\bf
V}\times\frac{d}{d{\bf p}}\right]_z\varphi_0\,.\end{equation}
where $v=\sqrt{v_x^2+v_y^2}$ and the integrand is called  the Berry connection (or curvature).
Everywhere, the summation over the band subscript is implied.

\begin{figure}[h]
\resizebox{.5\textwidth}{!}{\includegraphics{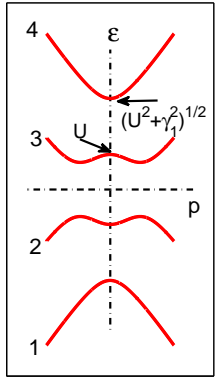}}
\caption{Band structure of bilayer graphene} \label{dis}
\end{figure}

Now let us calculate the $\delta$-phase for bilayer graphene. In the simplest  case, omitting $\gamma_3$ and $\gamma_4$, the effective Hamiltonian can be written  as 
\begin{equation}
H(\mathbf{p})=\left(
\begin{array}{cccc}
U \,    & q_{+} \,& \gamma_1    \, & 0\\
q_{-} \,& U     \, & 0\,& 0\\
\gamma_1    \,  &0 \,&-U  \,  &q_{-}\\
0 \,& 0 \,&q_{+} \,&-U
\end{array}%
\right) ,  \label{ham}
\end{equation}%
where the parameter $U$ describes the tunable gap as a result of the gate voltage and $\gamma_1$ is the interlayer nearest-neighbor hopping integral energy.
The constant velocity parameter $v$ is incorporated in the notation $q_{\pm}=vp_{\pm}$.
The band structure is shown in Fig. \ref{dis}. The minimal value of the upper energy $\varepsilon_4$ is $\sqrt{U^2+\gamma_1^2}$, the $\varepsilon_3$ band takes the maximal value $|U|$ at $q=0$.
Here, the orbit is the circle defined by Eq. (\ref{edm}), written in the  form
\begin{equation}
[(U+\varepsilon)^2-q^2][(U-\varepsilon)^2-q^2]-\gamma_1^2(\varepsilon^2-U^2)=0\,.
\label{disp}\end{equation}
The eigenfunction ${\mathbf \varphi_0}$ of the Hamiltonian
(\ref{ham}) can be taken as
\begin{equation}
{\mathbf \varphi_0}=\left(
\begin{array}{c}
(U-\varepsilon)[(\varepsilon+U)^2-q^2]\\
q_{-}[q^2-(\varepsilon+U)^2]\\
\gamma_1(U^2-\varepsilon^2)\\
\gamma_1q_{+}(U-\varepsilon)\end{array}\right) ,\label{fun}
\end{equation}
with the norm squared
\begin{eqnarray}
\varphi_0^*\varphi_0=[(\varepsilon+U)^2-q^2]^2[(\varepsilon-U)^2+q^2]\nonumber\\
+\gamma_1^2(\varepsilon-U)^2
[(\varepsilon+U)^2+q^2]\,.\label{norm}\end{eqnarray}
The derivatives for Eq. (\ref{on1}) are
calculated along the trajectory where the energy  $\varepsilon$ and, consequently, the trajectory radius $q$ are constant.

If the conditions $|U|<|\varepsilon|<\sqrt{U^2+\gamma_1^2}$  are fulfilled, Eq. (\ref{disp}) has only one solution for the radius squared  
$$q^2=U^2+\varepsilon^2+\sqrt{4U^2\varepsilon^2+(\varepsilon^2-U^2)\gamma_1^2}\,.$$ 
The matrix $V_y=\partial H/\partial p_y$ in Eq. (\ref{on1}) has  four nonzero elements, $V_{y}^{12}=V_{y}^{21}=V_{y}^{34}=V_{y}^{43}=-1$.

\begin{figure}[b]
\resizebox{.5\textwidth}{!}{\includegraphics{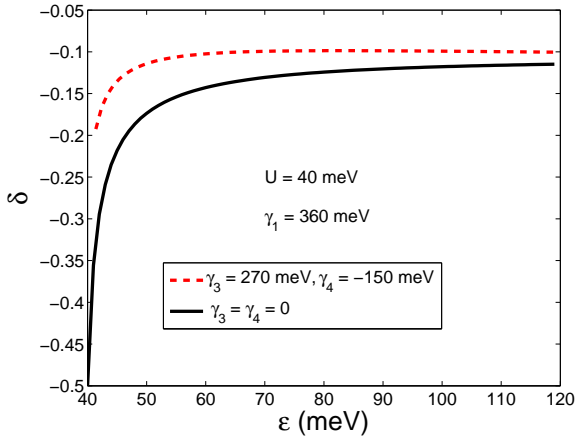}}
\caption{(Color online) Semiclassical phase vs  energy in the conduction band of bilayer graphene without trigonal warping (solid line) and with warping (dashed line).} \label{disg}
\end{figure} 
 Using Eqs. (\ref{disp}) and (\ref{fun}), we find
\begin{equation}\text{Im}\,\varphi_0^*V_y\frac{d\varphi_0}{dp_x}=
4U\varepsilon(U-\varepsilon)[(\varepsilon+U)^2-q^2]\,.\label{sver}\end{equation}
This
expression is  constant on the trajectory as well as $\varphi_0^*\varphi_0$, Eq. (\ref{norm}).
Therefore, in order to find $\delta$, Eq. (\ref{on1}), we have to
integrate along the trajectory
\[\oint
\frac{dp_x}{v_y}\,.\] This integral equals
$-dS(\varepsilon)/d\varepsilon$, where $S(\varepsilon)=\pi q^2$ is
the  cross-section area, Eq. (\ref{on}), with
\begin{equation}
\frac{dS(\varepsilon)}{d\varepsilon}=\pi\varepsilon\frac{2(q^2+U^2-\varepsilon^2)+\gamma_1^2}{q^2-U^2-\varepsilon^2}\,.
\label{dsec}\end{equation}

Now we have to substitute Eqs. (\ref{norm}) -- (\ref{dsec}) into Eq. (\ref{on1}).
Thus, we find  the Berry phase 
\begin{equation}
\delta(\varepsilon)=\frac{-\varepsilon U}{q^2-\varepsilon^2-U^2}=
\frac{-\varepsilon U}{\sqrt{4U^2\varepsilon^2+(\varepsilon^2-U^2)\gamma_1^2}}
\label{del1}\end{equation}
 shown in Fig. \ref{disg}, where $\delta$-phase of bilayer graphene with trigonal warping is also shown; the detailed calculations will be  published  elsewhere.
For the ungaped bilayer, $U=0$, the Berry phase $\delta(\varepsilon)=0$. The Berry phase
depends on the energy and $\delta=\mp 1/2$ at $\varepsilon =\pm U$. At the larger energy, $\varepsilon\gg U$, the Berry phase $\delta\rightarrow \mp U/\gamma_1$. 

\begin{figure}[b]
\resizebox{.5\textwidth}{!}{\includegraphics{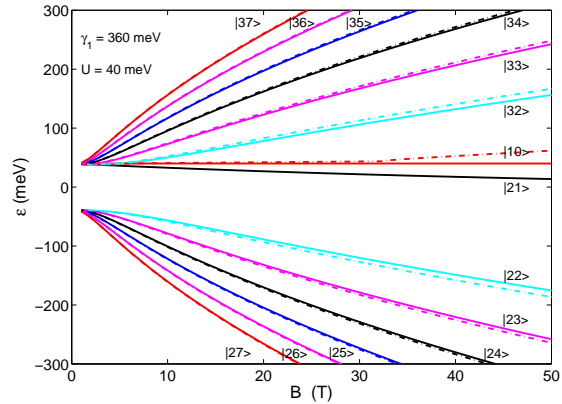}}
\caption{(Color online) Energy levels $\varepsilon_{sn}$ for the $K$ valley in magnetic fields for bilayer graphene within the perturbation approximation  (solid lines) and in the semiclassical approach (dashed-dotted lines); in the notation $|sn\rangle$, $n$ is the Landau number  and $s=1,2,3,4$ is the band number, only two nearest bands ($s=2,3$) are shown  at given $n$ from 0 to 7. There is only one level, $|10\rangle$, with $n=0$ and three levels ($s=1,2,3$) with $n=1$. The levels for the $K'$ valley can be obtained by mirror reflection with respect to the $\varepsilon=0$ axis.} \label{comp}
\end{figure}

Substituting Eq. (\ref{del1}) in the semiclassical quantization condition, Eq. (\ref{on}), and solving the equation obtained for $\varepsilon$, we get the energy levels as  functions of the magnetic field. We have to notice that the Landau numbers $n$ listed in Fig. \ref{comp} do not coincide with the numbers $n_c$ in the semiclassical condition (\ref{on}). The rigorous quantization shows that there are only one Landau level with $n=0$ and three Landau levels with $n=1$ \cite{Fa}. These levels are not correctly described within the semiclassical approach. However, for $n\geq 2$, there are levels in all four bands $s$ (two nearest bands with $s=2,3$ are shown in Fig. \ref{comp}). They correspond with the semiclassical number $n_c = n - 1$, and  the semiclassical levels for the larger $n$ are in excellent agreement with the levels obtained in the perturbation approximation.
 \section{Magneto-optics effects in graphene layers}  
 An important peculiarity of conductivities in  presens of  magnetic fields is an appearance  of the Hall component $\sigma_{xy}(\omega)$.
 The Hall conductivity violates the rotation symmetry  of graphene around the major axis. This implies  the rotation of the linear polarized electromagnetic wave, i. e.,  the Faraday and Kerr effects for transmitted and reflected waves, correspondingly.   First of all, the electron transitions are possible between the levels with the neighboring Landau numbers $n$ and various bands $s$, and therefore the resonance denominators $\Delta_{ss'n}=\varepsilon_{sn}-\varepsilon_{s', n+1}$ arise in the conductivity tensor.

Calculations  \cite{Fa} give the conductivities for graphite in the collisionless limit when the electron collision frequency $\Gamma$ is much less than the level splitting 
\begin{equation}
\begin{array}{c}
\left.\begin{array}{c} \sigma_{xx}(\omega)\\ i\sigma_{xy}(\omega)
\end{array}\right\}=i{\displaystyle\sigma_d
\frac{4\omega_B^2}{\pi^2}}
{\displaystyle\sum_{n,s,s'}\int\limits_0\limits^{\pi/2}dz\frac{\Delta f_{ss'n}}{\Delta{ss'n}}|d_{ss'n}|^2}\\ 
\times
\left[(\omega+i\Gamma
+\Delta_{ss'n})^{-1}\pm
(\omega+i\Gamma-\Delta_{ss'n})^{-1} \right]
\,,
\end{array}
\label{dc1}\end{equation}
where the integration is taken over the reduced Brillouin zone,  $0<z<\pi/2$. Such integration is absent for graphene and bilayer. 
Here $\Delta f_{ss'n}=f(\varepsilon_{s'n+1})-f(\varepsilon_{sn})$ is the difference of the Fermi functions and  
 \begin{equation}\begin{array}{c}
d_{ss'n}=C^2_{sn}C^{1}_{s'n+1}+C^{3}_{sn}C^{4}_{s'n+1}\nonumber\\
+(\tilde{\gamma}_4/\gamma_0)(C^1_{sn}C^{4}_{s'n+1}+C^{2}_{sn}C^{3}_{s'n+1})\end{array}\label{dip}\end{equation} 
is the dipole matrix element expressed in terms of wave functions (\ref{func}). These transitions are most intensive. They obey the
the selection rule
$$\Delta n=1\,,$$
and  will be referenced as the strong lines.
 The conductivity units  here
$$\sigma_d=\frac{e^2}{4\hbar d_0}$$
have a simple meaning, being  the  graphene universal conductivity    $e^2/4\hbar$ multiplied by the number  $1/d_0$ of layers within the  distance unit in the major axis direction.

Besides, we have to take  the renormalization of the dipole moments due to trigonal warping into account. This additional electron-photon vertex
results in weak lines with the selection rule
 $$\Delta n=2.$$ We get this contribution by substituting 
$$d_{ss'n}= (\tilde{\gamma}_3/\gamma_0)C^2_{sn}C^{4}_{s'n+2}$$
instead of the matrix element in
Eq. (\ref{dc1}) and replacing the subscript
$n+1\rightarrow n+2$.
We have to notice, that the $\gamma_4$ corrections give the linear (in small parameter $\gamma_4/\gamma_0$) contribution  to the conductivities at the main electron transitions with $\Delta n =1$. The $\gamma_3$ corrections are quadratic, however, they result in an appearance of  new resonant transitions with  $\Delta n =2$.

There are also small so-called vertex corrections  to the self-energy shown at the bottom of Fig. \ref{diag}. They  result from the  quartet of the coupled Landau levels, which interfere while  the selection rules $\Delta n=1$ and $\Delta n=2$ are allowed.

\subsection{Gapped bilayer graphene}

Graphene and bilayer graphene effect the transmission and the Faraday rotation in a linear order in the fine structure constant whereas the reflected light intensity is quadratic in $\alpha$. Therefore, let us discuss the
characteristics of the transmitted light through  bilayer graphene where the effects have a maximum value.  For this case, Eq. (\ref{dc1}) is valid without the integration over the $z$ momentum component. The conductivity units should be taken now as $\sigma_0=e^2/4\hbar$. In the approximation linear in conductivities, the transmission coefficient $T$ and the Faraday angle for the free standing bilayer  write  as  
\begin{equation}
1-T = \frac{4\pi}{c}\text{Re}\,\sigma_{xx},
\Theta_F= \frac{2\pi}{c}\text{Re}\,\sigma_{xy}\,.
\label{trbi}\end{equation}
 \begin{figure}[]
\resizebox{.5\textwidth}{!}{\includegraphics{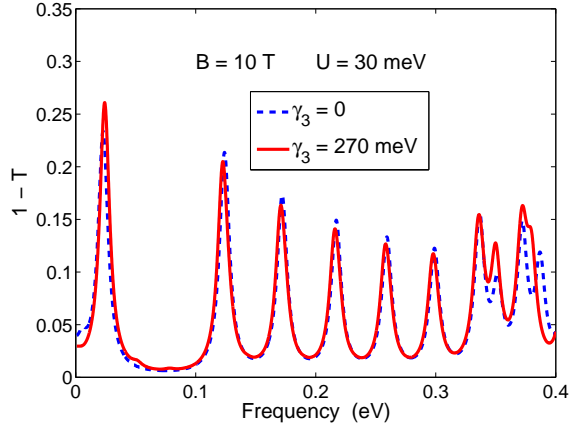}}
\caption{(Color online) Transmission spectra of gapped bilayer graphene without and with trigonal warping (dashed and solid lines, correspondingly) at  10 T and $U=30$ meV; the band parameters used are $v=1\times 10^8$ cm/s, $\gamma_1=360$ meV, $\gamma_4=-150$ meV, $\varepsilon_F=30$ meV, others are listed in Fig. The relaxation frequency is supposed as $\Gamma=5$ meV.}
\label{trabi}
\end{figure}

\begin{figure}[]
\resizebox{.5\textwidth}{!}{\includegraphics{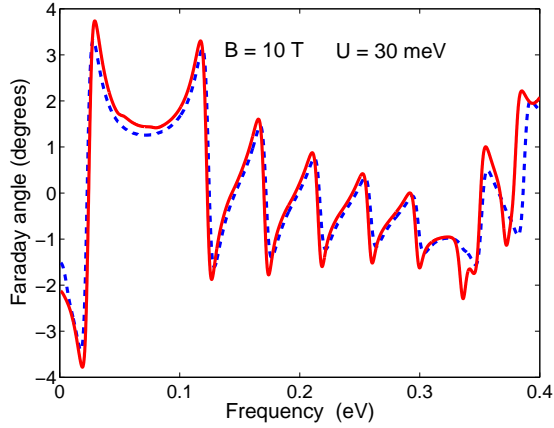}}
\caption{(Color online) Faraday rotation in gapped bilayer graphene; parameters used are the  same as in Fig. \ref{trabi}.}
\label{rabi}
\end{figure} 
 
 Results of  calculations are shown in Figs. \ref{trabi} and \ref{rabi}.
 The peaks in absorption, Fig. \ref{trabi} correspond to the  electron transitions. There is the series of seven lines in the 0.1--0.4 eV interval. They are doublets excited by the electron transitions of the type $|2n\rangle\rightarrow|3,n+1\rangle$ and  $|3n\rangle\rightarrow|2,n+1\rangle$ for $n$ from 2 to 8. Two weaker lines 
 at 350 and 380 meV are resulted from the  $|10\rangle\rightarrow|31\rangle$ and  $|21\rangle\rightarrow|42\rangle$
 transitions, correspondingly.  There is strongest line at 24 meV excited by the $|21\rangle\rightarrow|32\rangle$ transition.  All these lines obey the selection rule $\Delta n =1$. 

The very weak lines at  51 and 78 meV owe their appearance  to the $\Delta n=2$ transitions $|21\rangle\rightarrow|33\rangle$ and $|10\rangle\rightarrow|22\rangle$.

In general, the effect of the small constants $\gamma_3$ and $\gamma_4$ is seen more on the low levels $|10\rangle$  and $|21\rangle$.
 
The transition frequencies  in the Faraday rotation, Fig. \ref{trabi} are determined by the derivative maximum values. 

\begin{figure}[]
\resizebox{.52\textwidth}{!}{\includegraphics{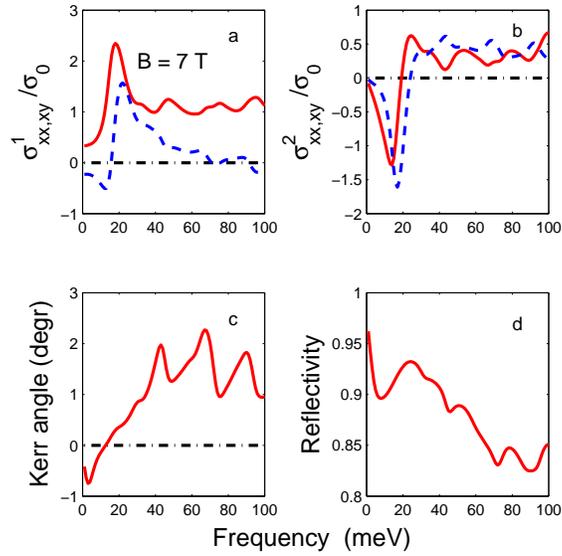}}
\caption{(Color online)  Real (a) and  imaginary (b) parts of the longitudinal (xx, solid line)  and Hall (xy, dashed line) dynamical conductivities calculated for one graphite layer in units of $\sigma_0=e^e/4\hbar$; Kerr angle (c) and reflectivity (d).
The magnetic field  $B=$ 7 T, the temperature T = 0.1 meV is less than the level broadening $\Gamma=3.5$ meV.}
\label{xx7}
\end{figure}

\subsection{Graphite}
Using the conductivities Eqs. (\ref{dc1}), one finds the complex bulk dielectric function $\varepsilon_{ij}=\delta_{ij}+4\pi i\sigma_{ij}/\omega$ and
the reflection coefficient and the Kerr rotation (see, e.g., \cite{LTK})
\begin{equation}
R=\frac{1}{2}(|r_+|^2+|r_-|^2), \Theta_K=\frac{1}{2}\arg(r_-r_+^*)\,,
\nonumber\end{equation}
where $r_{\pm}=(1-\sqrt{\varepsilon_{\pm}})/(1+\sqrt{\varepsilon_{\pm}})$  are the reflection Fresnel coefficients for two circular polarizations with $\varepsilon_{\pm}=\varepsilon_{xx}\pm \varepsilon_{xy}$. 
\begin{figure}[]
\resizebox{.5\textwidth}{!}{\includegraphics{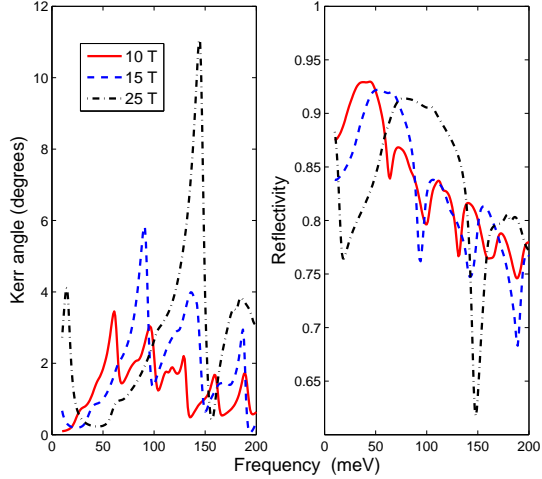}}
\caption{(Color online) Kerr angle and reflectivity at  10, 15, and 25 T.}
\label{kerrang1}
\end{figure} 

The parameters of Eq. (\ref{hamg}) used in the calculations are listed in Table \ref{tb1} (see also Ref. \cite{BCP}). The  hopping integrals $\gamma_0$ to
$\gamma_3$ are close to the values determined in observations of the semiclassical ShdH effect. The Fermi energy equal to $\varepsilon_F=-4.1$ meV  agrees at the zero magnetic field   with the measurements of the extremal Fermi-surface cross sections and the masses of holes and electrons. 
Connections with the notation for  similar parameters of the SWMC model are given in the "SWMC"\, line.
  The values of parameters $\gamma_4$, $\gamma_5$, and $\Delta$ determined in various experiments are very different; we use $\gamma_5$ and $\Delta$  obtained by Doezema et al \cite{DDS} (given in Table \ref{tb1} in the "SWMC"\, notations) and take  the approaching value for $\gamma_4$. In the quantum limit, when electrons and holes occupy only  $|10\rangle$ and $|21\rangle$ levels, the Fermi energy must cross these close levels at the middle of the $KH$ line. It means that the Fermi level becomes higher at such the magnetic fields   taking the value $\varepsilon _F\approx -1$ meV.

The results of calculations are represented in Figs. \ref{xx7}-\ref{kerrang1}. Let us emphasize that
 the imaginary part of the dynamical conductivity is of the order of the real part.

One can see in Fig.  \ref{xx7} (a), that  the  averaged longitudinal conductivity calculated per one
graphite layer tends   to the  graphene universal conductance.
The main contribution in the sharp 16-meV line is resulted from
the electron $|21\rangle\rightarrow|32\rangle $ transition (15 meV) about the $K$ point (see Fig. \ref{d7strf}) where the $|32\rangle $ level coincides with the Fermi level (within an accuracy of the width $\Gamma$ or temperature $T$). 
 Then, the transitions $|22\rangle\rightarrow|21\rangle $ produce 
the broad band. The low-frequency side of the band (23 meV, at the intersection of the  $|21\rangle $ level with the  Fermi level) contributes into  the  16-meV line. 
In the same 16-meV line, the transitions $|32\rangle\rightarrow|33\rangle $ can contribute as well if the band $|32\rangle$ contains the electrons.

The next doublet at 43 meV  arises  from the transitions $|23\rangle\rightarrow|32\rangle $  and $|22\rangle\rightarrow|33\rangle$  at the $K$ point.
The 68- meV doublet  appears   as the splitting of the  $|24\rangle\rightarrow|33\rangle $ (65 meV) and $|23\rangle\rightarrow|34\rangle$ (69 meV) transitions due to the electron-hole asymmetry at the $K$ point  of the Brillouin zone.   
 
 The 89-meV line is more complicated. First,  there are the electron transitions  $|24\rangle\rightarrow|35\rangle\,$ (89 veV) and $|25\rangle\rightarrow|34\rangle $ (90 meV) near the $K$ point. 
  Besides,   the transitions $|11\rangle\rightarrow|10\rangle $ (95 meV) near the $H$ point make  a contribution as well. All these lines  obeying the selection rule $\Delta n=1$ are strong. There are two weak lines in the frequency range. One ($|24\rangle\rightarrow|32\rangle $) is seen at 55 meV 
  as a shoulder  on the theoretical curve. Another, at 31 meV, results from the transitions $|10\rangle\rightarrow|32\rangle $ near the $K$ point.
 
The positions of the lines  for  fields in the range of 10 -- 30~ T agree   with observations of Refs. \cite{OFS,CBN}.
   
The optical Hall conductivity $\sigma_{xy}(\omega)$ in the ac regime is shown in Figs.   \ref{xx7} (a) and \ref{xx7} (b). 
The conductivities 
$\sigma_{xx}(\omega)$ and $\sigma_{xy}(\omega)$ allow  calculating the Kerr rotation and the reflectivity  as functions of frequency [see  Figs. \ref{xx7} (c) and \ref{xx7} (d)]. 
It is evident that the interpretation of the Kerr rotation governed by
the conductivity $\sigma_{xy}(\omega)$  is much more complicated
  in comparison with the longitudinal conductivity.  
 The Kerr angle and reflectivity shown in Fig. \ref{kerrang1} for the different magnetic fields demonstrate the strong field dependence of the magneto-optic phenomena.

 \section{Summary and conclusions}

In conclusions, we have evaluated the perturbation theory for the matrix Hamiltonian, which permits to calculate the corrections to  eigenvalues resulting from the small matrix elements particularly  from the trigonal warping. The trigonal warping in graphite can be considered within the perturbation theory at  strong magnetic fields larger than 1 T approximately. For weak magnetic fields, when the Fermi energy  much larger than the cyclotron frequency, the semiclassical quantization with the Berry phase included can be applied. We have found that the main electron transitions obey the selection rule $\Delta n=1$ for the Landau number $n$, however the $\Delta n=2$ transitions due to the trigonal warping with the small probability are also essential. In graphite, the electron transitions  at the $K$, $H$ points as well as at  intersections of the Landau levels with the Fermi level make  contributions into conductivity. The good agreement between the calculations and the measured  Kerr rotation and reflectivity in graphite in the quantizing magnetic fields is achieved. The SWMC  parameters are used in the fit  taking  their values from the previous dHvA measurements  and increasing the Fermi energy value for the case of the strong magnetic fields. 

\acknowledgments
The author acknowledges useful discussions with A. Kuzmenko and J. Levallois.
This work was supported by the Russian Foundation for Basic
Research (grant No. 10-02-00193-a) and the SCOPES grant IZ73Z0$\_$128026.


\end{document}